# Degradation by Exposure of Co-Evaporated $CH_3NH_3PbI_3$ Thin Films

*Youzhen Li, Xuemei Xu, Chenggong Wang, Congcong Wang, Fangyan Xie, Yongli Gao*[*]

Dr. *Youzhen Li*, Dr. *Xuemei Xu*, School of Physics and Electronics, Central South University, Changsha, Hunan, 410083, P.R. China
Dr. *Chenggong Wang, Congcong Wang*, Prof. *Yongli Gao*, Department of Physics and Astronomy, University of Rochester, Rochester, NY 14627, USA
ygao@pas.rochester.edu
Dr. *Fangyan Xie*, Instrumental Analysis Center, Sun Yat-Sen University, Guangzhou, 510275, P.R. China

During the past several years, organometal halide perovskites have attracted much attention as light-harvesting materials for solar cells.[1-5] The advantages of this kind of solar cells are wide light-absorption spectra, high efficiencies and low costs.[6-9] Perovskite solar cell was first reported by Miyasaka and co-workers in 2009, with the power conversion efficiency (PCE) of 3.8%.[10] The first solid-state perovskite solar cell with a PCE of 9.7% was reported in 2012 by H.S. Kim and co-workers.[2] Ball et al. found that planar heterojunction perovskite solar cell could get around 5% efficiency.[11] This means that perovskite solar cells can be fabricated in an easier way with lower cost and higher produce efficiency. The PCE was quickly improved to 15.4%,[12] 19.3%[13] and 20.1%[14] in the following years. This is very close to the PCE of the best silicon solar cell. Many new kinds of perovskite solar cells with planar heterojunction structure have been fabricated in the last 2 years, such as lead free perovskite [15, 16] and hole-conductor-free perovskite.[17] Solution spin cast and thermal evaporation are two main methods to get perovskite films. However, the surface of a spin cast film is usually not uniform nor stoichiometric, not particularly suitable for surface sensitive investigations.[18-24] Fabrication by evaporation was first reported by Snaith's group.[12] They got perovskite films with a much uniform surface, and obtained devices of higher efficiency and open circuit voltage than that of spin cast one.

A critical issue of the perovskite solar cells is the stability. The performance degraded quickly when the solar cells were put in ambient, which can be a show stopper for perovskite solar cells to reach the market. Grätzel and co-workers reported that perovskite solar cell could be fabricated under controlled atmospheric conditions with a humidity <1%.[25] Yang and co-workers reported that perovskite solar cell prepared by $PbCl_2$ and $CH_3NH_3I$(MAI) in controlled moisture environment could get a good crystal structure and the PCE of 17.1%.[26] The reports all mentioned that the performance of perovskite solar cell was sensitive to moisture. Qiu and co-workers observed the degradation process with UV-Vis spectroscopy and X-ray diffraction (XRD).[27] They proposed that the degradation progressed in two steps. First, $CH_3NH_3PbI_3$ decomposed into $CH_3NH_3I$ solution and $PbI_2$ in the presence of $H_2O$. The $CH_3NH_3I$ then decomposed into $CH_3NH_2$ and HI, and HI might react with $O_2$ or decompose under illumination into $H_2$ and $I_2$. They suggested that $H_2O$ acted as a catalyst during the process.[27] The interpretation is interesting and the detailed reaction process and mechanism of the perovskite with the environment need further investigation.

In this manuscript, we report our investigations on the degradation mechanisms of $CH_3NH_3PbI_3$ using XPS and XRD. The $CH_3NH_3PbI_3$ films were fabricated by co-evaporation of $PbI_2$ and $CH_3NH_3I$, and excellent atomic ratio and crystal structure were observed. The films were then exposed to oxygen, air or $H_2O$ with pre-selected exposures, and the spectroscopic change was closely monitored to reveal the evolution of each element. We found that the $CH_3NH_3PbI_3$ film was not sensitive to oxygen and dry air. It was, however, quite sensitive to moisture. Before $10^{10}$ L $H_2O$ exposure, the film is stable and $H_2O$ acts as an n-type dopant. After about $2 \times 10^{10}$ L of exposure, it starts to decompose, characterized by the complete removal of N and part of I. The remnants of the film on the surface are crystalline $PbI_2$, hydrocarbon complex and O contamination.

The XPS survey of as deposited $CH_3NH_3PbI_3$ film and the core level spectra are shown in Figure1(a) and (b), respectively. No Au substrate signal was detected, indicating that the surface was uniform. The C 1s peak of the $CH_3NH_3PbI_3$ film consists of the carbon of $CH_3NH_3PbI_3$ film at 286.02 eV and a small amount (~1/8) of amorphous C contamination at 284.52 eV. Some of the amorphous C contamination may start from the Au substrate contamination, as we see it gradually reduce as the film thickness increases. After peak fitting with commercial Origin (see Figure1(b)) and taking into account the instrumental atomic sensitivity factor (ASF) as well as removing the amorphous C contamination, we found that the atomic ratio was C: N: Pb: I = 1.34:1.07:1.00:3.15, very close to the ideal $CH_3NH_3PbI_3$. The inset of Figure 1(a) shows the valence band maximum (VBM) and vacuum level cut-off of the as evaporated $CH_3NH_3PbI_3$ film. We got the VBM and work function (WF) that were 0.85 eV and 4.86 eV, respectively. The ionization potential of $CH_3NH_3PbI_3$ can be deduced to be 5.71 eV, substantially lower than those obtained from spin cast films whose stoichiometry are known to deviate from ideal.[19,28] The band gap of $CH_3NH_3PbI_3$ was reported to be around 1.55 eV,[28,29] so the prepared film was a slightly n-type film.

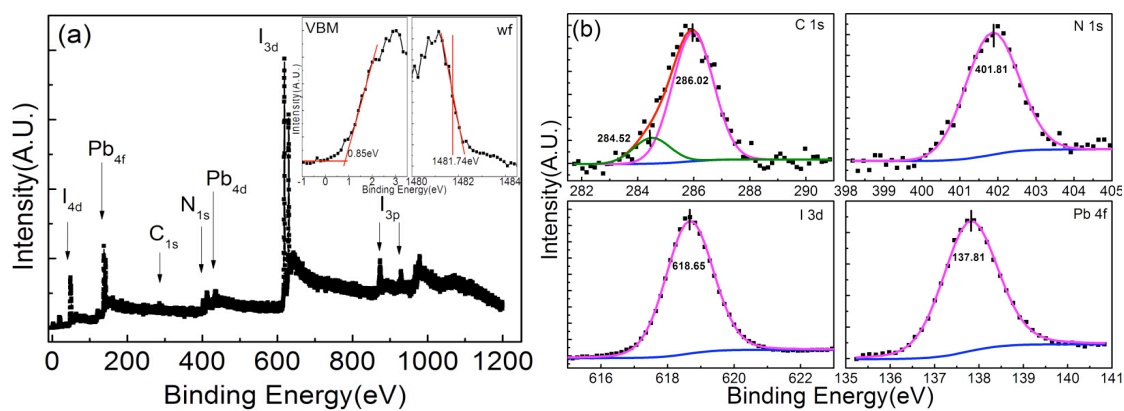

**Figure 1.** ( (a) XPS survey of as deposited $CH_3NH_3PbI_3$ film and(b) the core level spectra of the elements. The fitting curves are also shown. The inset of the left panel shows the valence band and vacuum level cut-off region.)

The XPS core level spectra of the $CH_3NH_3PbI_3$ film before and after exposure to oxygen are shown in Figure 2. No obvious change could be observed. It indicated that the film was not sensitive to oxygen. Even after $10^{13}$ L exposure, no O atoms can be found. The binding energy of C 1s, I 3d, and Pb 4f (Figure S1) have no discernable change, too. The slight VBM shift from 0.85 eV (0 L exposure) to 0.67 eV ($10^{13}$ L exposure) shows that the oxygen may act as a p-type dopant in the film without significant change of the film composition and structure.

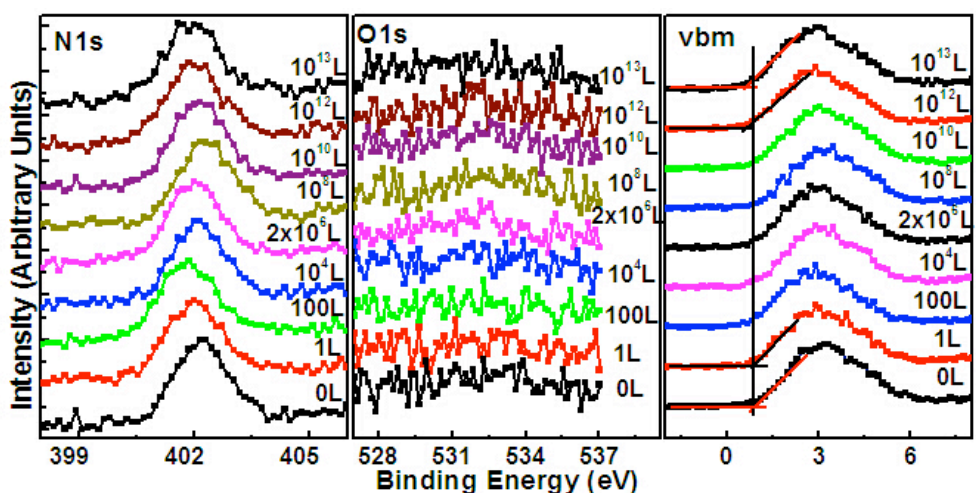

**Figure 2.( O 1s, N 1s, and VBM region XPS spectra of $CH_3NH_3PbI_3$ film during oxygen exposure. )**

Although all the atomic core level energies show no obvious change after the oxygen exposure, a closer examination reveals that the atomic ratio has changed slightly. Figure 3 shows the atomic ratio change of the film during the exposure. C 1s(A) (286.02 eV, carbon of the perovskite) and C 1s(B) (284.5 eV, amorphous carbon contamination peak) are the C 1s peaks of the film obtained from peak fitting (see Figure 1(b)). We can see that after $10^{13}$ L oxygen exposure, the atomic ratio of I 3d, N 1s and C 1s (A) decrease by about 0.32, 0.17 and 0.24, respectively. It seemed that about 20% of $CH_3NH_3I$ went away from the surface. The iodine escaped more as it might have sublimed. As the valence structure and the bulk part of the atomic ratio remained unchanged after the exposure, the sample was still mostly $CH_3NH_3PbI_3$. The intensity change may not be due to the oxygen exposure but to the stability of the sample in vacuum. Similar stability is also observed when the perovskite sample is exposed to air.

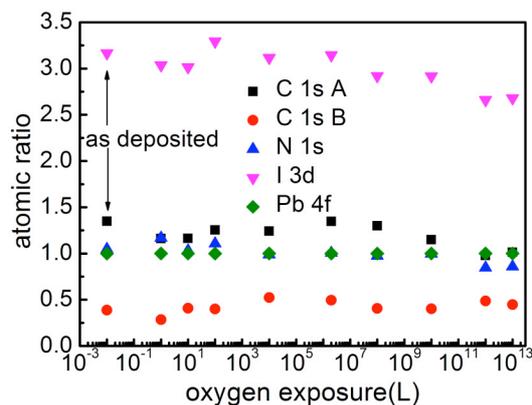

**Figure 3. The atomic ratio before and after oxygen exposure**

The exposure to air was done in the evaporation chamber. After exposure from 1 L to $10^{12}$ L, no obviously change could be observed (Figure S2) except for some increase of carbon contamination at ~284.5 eV from 15% to 40% of the C 1s peak. As the inner surface of the evaporation chamber is large and the stainless wall has a strong affinity to moisture, the limited moisture leaked into the chamber with air may partly be adsorbed by the chamber surface. As a result, the air was dry and the sample showed no sensitivity to the exposure. The XPS spectra of C 1s, N 1s, I 3d, O 1s, Pb 4f and VBM of the film before and after exposure to air in ambient are shown in Figure 4. For visual clarity, we have normalized the spectra to the same height except those of N 1s, as the latter disappears as the exposure progresses. After exposure of 60 min., the perovskite C 1s(A) peak shifted from 286.02 eV to 286.21 eV, and the intensity decreased about 50%. The amorphous C 1s(B) peak, on the other hand, increased by 110%. This indicates some contamination of amorphous carbon by the ambient exposure, as well as by the C leftover by the decomposition of the perovskite. After further exposure for 90 and 150 min., the perovskite C 1s(A) at 286.2 eV was all gone, and only amorphous C 1s(B) at 284.3 eV was left. The intensity of N 1s also decreased about 41% after the exposure of 60 min., and no N 1s could be detected after 90 min. of exposure. This indicated that the $CH_3NH_3PbI_3$ film had decomposed and the structure was altered. The intensity of Pb 4f almost has no change after exposure. The intensity of I 3d decreased only about 15% after 60 min. exposure in ambient, and the ratio of I to Pb changed from 3.15 to 2.5, consistent to the decomposition and partial removal of ~40% of $CH_3NH_3I$.

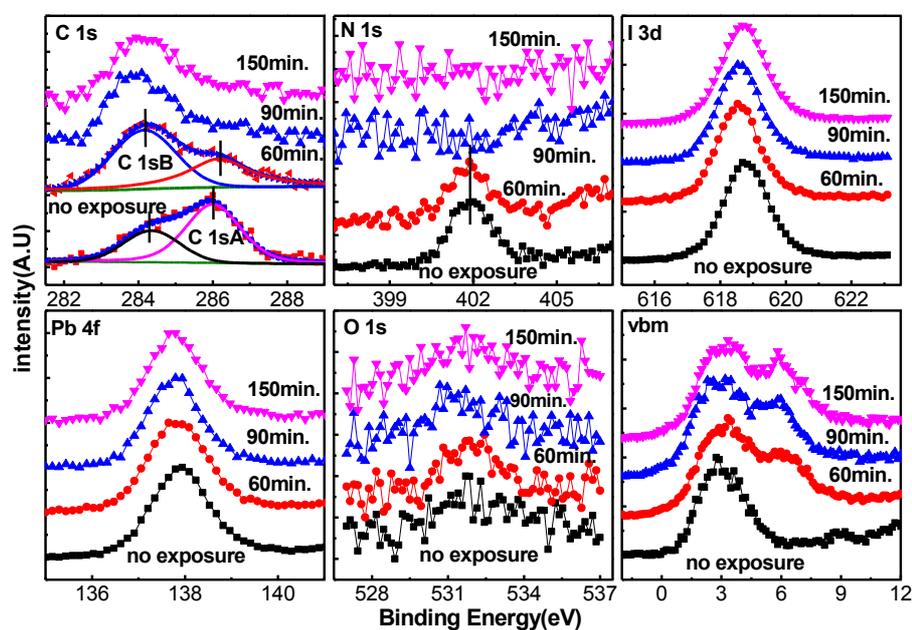

**Figure 4. XPS spectra during ambient exposure. The results point to the gradual removal of N after ambient exposure.**

After 150 min. exposure, the I 3d intensity decreased by about 30%, and the ratio of I to Pb changed to 2.2. It seemed that while $PbI_2$ and C left on the surface, N and part of I escaped the sample as gases. An interesting observation is that the intensity of O 1s is only slightly increased, suggesting that $CH_3NH_3PbI_3$ is not sensitive to oxygen even with some moisture. From the VBM curve, we can see that after 60 min. ambient exposure, a wide peak around 6.5 eV appeared, and became sharper with the time elapsed. This indicates the existence of $H_2O$ on the surface. The results confirm that $H_2O$ plays an important role in the degradation of $CH_3NH_3PbI_3$ film.

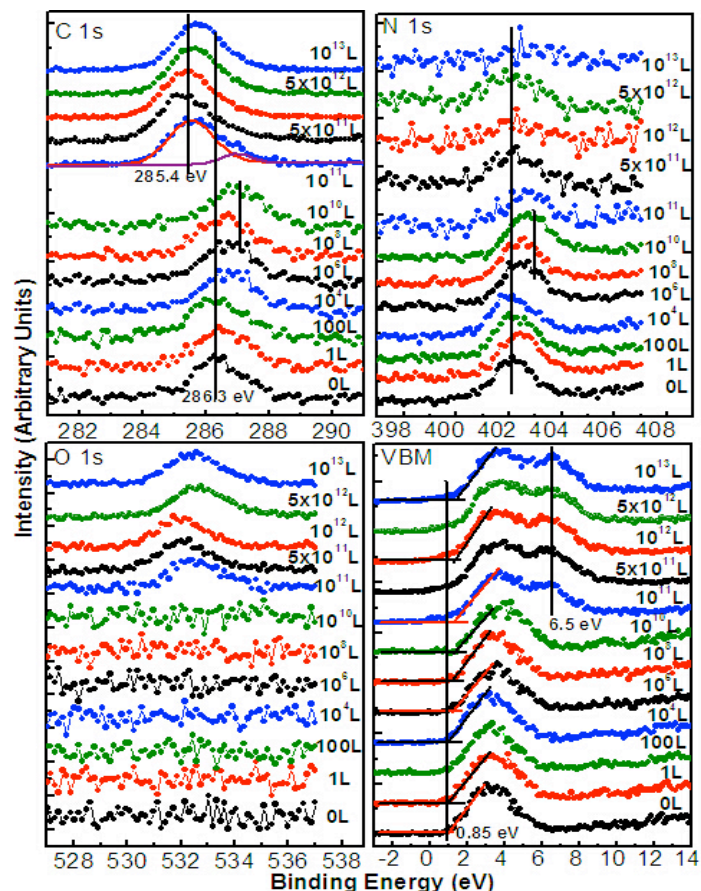

**Figure 5.( XPS spectra before and after $H_2O$ exposure. There's an initial rigid shift indicating n-doping till $10^{10}$ L exposure. Higher exposures introduce spectral changes indicating the decomposition of the perovskite and $H_2O$ adsorption.)**

The exposure studies in $O_2$ and air suggest that the perovskite is sensitive to moisture but $O_2$ or $N_2$. To get more insight of the process, we took the third sample to expose to $H_2O$ in the evaporation chamber in well controlled exposure sequence and well isolated from other possible factors. In Figure 5 are the XPS spectra of the C 1s, N 1s, O 1s and VBM before and after $H_2O$ exposure. The I 3d and Pb 4f spectra are not shown here as they have no significant change except some intensity reduction. Again the C 1s spectra are normalized to the same height for visual clarity, while those of N 1s and O 1s are not normalized because their late disappearance (appearance) at higher exposures. It is obvious that before $1 \times 10^{10}$ L exposure, there is no significant change in the spectral shape but a rigid shift of 0.65 eV to the higher BE. The rigid shift indicates that at this stage $H_2O$ acts as an n-dopant that moves the Fermi level of the perovskite from 0.85 to 1.41 eV from the VBM. Given the band gap of 1.55 eV for the perovskite, the shift puts the Fermi level very close to the bottom of the conduction band, i.e., the perovskite has been n-doped by $H_2O$. The integrity of the perovskite remains unaltered at this stage.

After $10^{11}$ L exposure, dramatic change occurred. The C 1s changed mainly from that of the perovskite(286.3 eV) to hydrocarbon complex (285.4 eV). The carbon of $CH_3NH_3PbI_3$ became very weak and was almost gone after $5\times10^{11}$ L $H_2O$ exposure.

Other spectral features also support the notion that decomposition of the perovskite occurs between $10^{10}$ to $10^{11}$ L. The N 1s peak became very weak, and no N can be found on the surface after $10^{12}$ L exposure. O 1s peak also appeared after $10^{11}$ L exposure, the intensity of O 1s is much higher than that of ambient exposure. We already knew that the $CH_3NH_3PbI_3$ film was not sensitive to oxygen. The high intensity of O must come from the adsorption of $H_2O$. The valence band spectra also show the same trend of O increase. After $10^{10}$ L $H_2O$ exposure, the VBM shift from 0.85 eV to 1.41 eV, and continue to ~1.50 eV after $10^{13}$ L exposure. The core level and VB have approximately the same amount of shift within the measurement uncertainty. Furthermore, a new wide peak centered at about 6.5 eV appeared after $10^{11}$ L exposure. All the spectra are consistent with the picture that the perovskite is n-doped till saturation at ~$10^{10}$ L exposure, after which a chemical reaction takes place that destroys the perovskite, leaving the surface with $PbI_2$, hydrocarbon complex , and O contamination most likely from the adsorption of $H_2O$.

Further insight can be obtained by observing the XPS peak shift and the change in the atomic ratio. Figure 6(a) shows the peak shifts during the $H_2O$ exposure. Before $10^{10}$ L exposure, all the core levels and VBM shift together. The rigid shift is indicative of doping, or change of the Fermi level within the band gap. The initial shift for <$10^4$ L exposure moves the peaks slightly to lower BE, and the surface appears slightly more p-type. The effect may be doping or change of surface screening.[30] It then shifts to higher binding energy till $10^{10}$ L, at which the exposure makes it strongly n-type. After $10^{11}$ exposure, all of peaks have sudden and discorded shifts towards low binding energy, a strong signature of chemical reactions. The results indicate that the film keeps chemically stable before $10^{10}$ L exposure. After $10^{10}$ L exposure, the drastic change shows that the film begins to decompose. The VBM shift makes it clearer. Before $10^6$ L exposure, the shift are all less than 0.3 eV, the film is stable, after $10^{10}$ L exposure, the VBM shifts 0.56 eV and gets to about 1.41 eV. The band gap of $CH_3NH_3PbI_3$ is reported to be 1.55 eV. The VBM shift means that before the film decomposition, $H_2O$ just as a dopant and make the perovskite more n-type.

The change of the atomic ratios after $H_2O$ exposure was shown in Figure 6(b). The obviously change also happened between $10^{10}$L and $10^{11}$L exposure, which indicated the film began to decompose. N went away quickly, and the intensity of I 3d decreased. At the end of the exposure, the atomic ratio of I to Pb decreased to 2.1, and the intensity of N 1s decreased to 0. It also indicated that the film decomposed, N and I left the surface, and only $PbI_2$, C and O compounds remained on the sample surface. The dramatic increase of C 1s intensity is quite remarkable. Contamination from the residual hydrocarbon in the evaporation chamber is unlikely as it should be linearly proportional to the time of exposure instead of a sudden increase once the perovskite starts to decompose. A more reasonable interpretation is that the hydrocarbon is expelled by the formation of $PbI_2$ and segregates to the surface, resulting in a strong signal from the surface sensitive XPS.

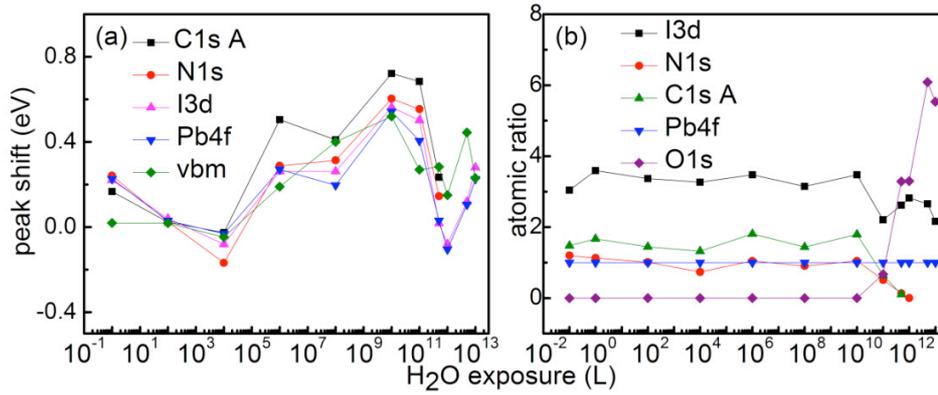

Figure 6.( (a) XPS peak shift and (b) atomic ratio change from the core level intensity after H₂O exposure.  The rigid shift for exposure =< 10¹⁰ L indicates the initial doping.  The decomposition at higher exposures is evident from the change of energy level shift and intensity.)

From the above discussion, we can conclude that after the H₂O exposure, the film decomposed as the following:

$$CH_3NH_3PbI_3 \xrightarrow{H_2O} (-CH_2-) + NH_3(g) + HI(g) + PbI_2 \qquad (1)$$

This bears resemblance to the water catalyzed reaction model proposed in Ref. 27, with the important difference that there are gaseous species that subsequently escape from the sample, leaving PbI₂ and hydrocarbon complex behind.  （-CH2-）represents the residual hydrocarbon complex, most likely polyethylene-like (-CH₂-CH₂-) from the corresponding C 1s energy of 285.4 eV.[31]

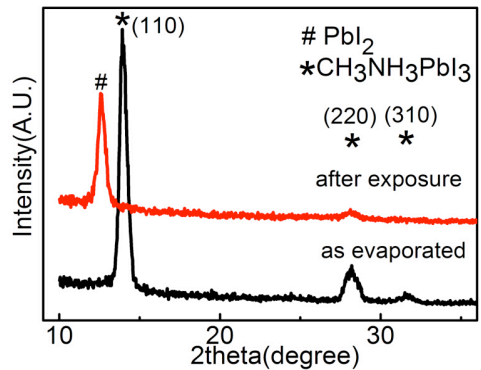

Figure 7.( XRD scan of CH₃NH₃PbI₃ film before and after ambient exposure.  The decomposition of the perovskite to PbI₂ by the exposure is evident.  The (220) peak of the perovskite is still discernable after the exposure, indicating that deep inside the sample remains intact.)

Further support of our interpretation can be obtained by the structure investigation. Figure 7 shows the XRD spectrum of the $CH_3NH_3PbI_3$ film before and after ambient exposure. It is obvious that the as-evaporated film has a strong 110 diffraction peak with $2\theta =13.9°$, which indicates that the $CH_3NH_3PbI_3$ film had a good crystallinity. $CH_3NH_3PbI_3$ films prepared by spin cast must be annealed to get the crystal structure. For evaporated ones, annealing is not necessary provided the evaporation rate is kept moderate. After air exposure, the film degraded and the perovskite (110) diffraction peak is disappeared, and the $PbI_2$ diffraction at $12.7°$ became the dominant feature. It is interesting to note that while the (110) feature is completely gone after the exposure, the (220) peak is still discernable. This is because of the larger angles allows the x-ray to penetrate deeper into the film, whose integrity remains intact deeper inside.

It turns out that our $H_2O$ and air exposure experiments provide strong support to each other and allow us to quantify the criterion of decomposition by moisture. As $CH_3NH_3PbI_3$ film is sensitive to moisture and insensitive to oxygen, we can calculate the damage threshold by moisture from the ambient exposure. We found that the film began to decompose after about 60 min. in ambient. The relative humidity of the ambient in our lab was 41%. The saturated vapor pressure of water at room temperature (21℃) is about 18.650 torr.[32] The 60 min. ambient exposure amount to $0.41 \times 60 \times 60 \times 18.650 = 2.75 \times 10^{10}$ L, consistent to the $H_2O$ exposure measurements that put the threshold between $10^{10}$ and $10^{11}$ L. We can conclude that before about $2 \times 10^{10}$ L $H_2O$ exposure, the $H_2O$ acts as an n-dopant. Beyond that, it starts to decompose the perovskite and the structure is quickly destroyed. Shown in Figure 8 is the schematic of the two stages of the interaction process during $H_2O$ exposure of the perovskite.

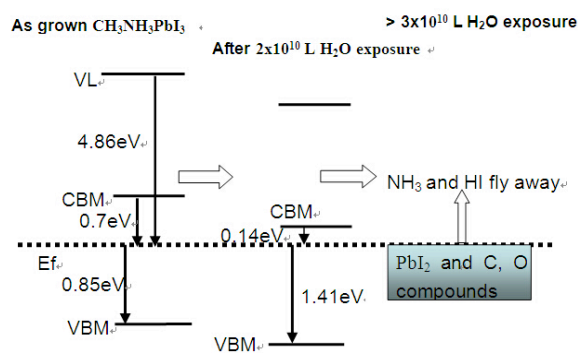

**Figure 8.( Schematics of perovskite decomposition by $H_2O$ exposure. The process of exposure can be separated into the doping stage and destruction stage.)**

Niu *et a*l. proposed that with the existence of $H_2O$, $CH_3NH_3PbI_3$ should decompose into $PbI_2$, HI and $CH_3NH_2$, and then turn into $I_2$, $H_2$ and $H_2O$ with the existence of $O_2$ or light.[27] Our observation supports the notion of water catalyzed decomposition, and our model presented in Equation(1), hereby termed water catalyzed vaporization (WCAV), provides further insight on the process that decomposition and vaporization of the gaseous species deplete the sample of N and I, and leave the surface of the sample with $PbI_2$, hydrocarbon complex, and $H_2O$ adsorption. This is significant as it shows the irreversibility of the degradation should the gaseous species are allowed to escape.

We have investigated the degradation mechanisms of $CH_3NH_3PbI_3$ using XPS and XRD. $CH_3NH_3PbI_3$ films with the right atomic ratio and crystal structure were successfully fabricated by co-evaporation. We find that the $CH_3NH_3PbI_3$ film is not sensitive to oxygen. It is quite sensitive to moisture and the interactions can be characterized in two stages. The first stage is for $H_2O$ exposure less than about $2 \times 10^{10}$ L, in which $H_2O$ acts as an n-dopant and the integrity of the perovskite remains intact. The first stage ends when the surface of the perovskite becomes highly n-doped. Higher exposure leads to the second stage when the perovskite quickly decomposes. The decomposition is characterized by the complete removal of N and part of I. The remnants of the film on the surface are crystalline $PbI_2$, hydrocarbon complex and O, the latter of which most likely from the adsorption of $H_2O$ on the surface. A model of water catalyzed vaporization (WCAV) is proposed based on our observations.

**Experimental Section**

*X-ray photoelectron spectroscopy*: The XPS measurements were performed in a modified Surface Science Laboratories' SSX-100 and was used to confirm the C , N, I and Pb content of the perovskite film. The monochromatic Al Kα radiation (1486.6eV) was used to excite photoelectrons under ultra-vacuum.

*X-ray diffraction*: The structure of the films was characterized by (XRD) collected with a Philips APD diffractometer, equipped with a Cu Kα X-ray tube operated at 40kV and 30 mA using a step size of 0.030 degrees and a time per step of 1.0 s. Experimental fitting of the X-ray data was carried out from 10-70°, $2\theta$ was fixed at an angle of 1 degree.

*Sample fabrication:* The substrates were Au covered silicon wafers. They were cleaned by using methanol in ultrason before loaded into the evaporation chamber. PbI$_2$ (Shanghai Zhenpin Com., 99%) and CH$_3$NH$_3$I (Wuhan Crystal Solar Cell Technology Com.) powder were put into tungsten boats separately. Each boat was attached a thermal couple that was tightened near the center of the boat to get the accurate temperature. The film atomic ratio was calculated by using the fitting peak areas divided by the atomic sensitive factors (ASF) of the instrument. The film thickness (mass equivalent thickness) was monitored by a quartz crystal microbalance. All the samples were fabricated with the same evaporation parameters.

*Exposure experiments details:* The MAI and PbI$_2$ growth rate were both kept at ~1 Å/min. For the oxygen exposure, ultrapure carried grade oxygen (UN1072) from Airgas Inc. was used in controlled steps from 1 L to $10^{13}$ L. The exposures up to $10^4$ L were performed in the spectrometer chamber. The rest exposures were performed in the evaporation chamber. After exposure, samples were transferred into the spectrum chamber for xps measurements, marks were made on the sample holder to make sure that each time the measurement was at the same position. The air exposure was performed in the evaporation chamber from 1 L to $10^{12}$ L. The ambient exposure was done at room temperature 21℃ and relative humidity 41%. The $H_2O$ exposure was administered with a clean glass tube half filled with deionized water, connected to the evaporation chamber through a micro-leak valve. The tube was pumped repeatedly to ensure that it was filled by pure $H_2O$ vapor. The XPS samples were 100 Å thick. The XRD sample was ~600 Å thick. It was inserted into a $N_2$ filled bag and brought to the XRD instrument within 10 min. out of the evaporation chamber to minimize ambient exposure. All the measurements were performed at room temperature.

**Acknowledgements**

The authors would like to acknowledge the support of the National Science Foundation Grant No. CBET-1437656 and DMR-1303742. YL, XX and FX acknowledge the support of China Scholarship Council.

[1]. G. Hodes and D. Cahen, *Nature Photonics* **2014**, *8*, 87.

[2]. H. S. Kim, C. R. Lee, J. H. Im, K. B. Lee, T. Moehl, A. Marchioro, S. J. Moon, R. Humphry-Baker, J. H. Yum, and J. E. Moser, *Scientific reports* **2012**, *2*,509.

[3]. M. M. Lee, J. Teuscher, T. Miyasaka, T. N. Murakami, and H. J. Snaith, *Science,* **2012,***338*, 643.

[4]. D.Liuand, T. L. Kelly, Nature photonics, **2013,** *8*,133.

[5]. S. D. Stranks, G. E. Eperon, G. Grancini, C. Menelaou, M. J. P. Alcocer, T. Leijtens, L. M. Herz, A. Petrozza, and H. J. Snaith, *Science,* **2013,***342* ,341,.

[6].T. Baikie,Y. N. Fang, J. M. Kadro, M. Schreyer, F. X. Wei, S. G. Mhaisalkar, M. Graetzel, and T. J. White, *J. Mater.Chem.* **2013**,*A1(18)*, 5628.

[7]. B. Conings, L. Baeten, C. D. Dobbelaere, J. D'Haen, J. Manca and H. G. Boyen, *Adv. Mater*. **2014**,*26(13)*, 2041,.

[8]. C.Wehrenfennig, G. E. Eperon, M. B. Johnston, H. J. Snaith and L. M. Herz, *Adv. Mater.* 2014,*26(10)*, 1584.

[9]. G. Xing, N. Mathews, S. S. Lim, N. Yantara, X. Liu, D. Sabba, M. Grätzel, S. Mhaisalkar, and T. C. Sum, *Nature materials*, **2014,** *13 (5)*, 476

[10]. A. Kojima,K. Teshima,Y. Shirai, and T. Miyasaka, *J.Am. Chem. Soc.* **2009**,*131*, 6050.

[11]. J. M. Ball, M. M. Lee, A. Hey, H. J. Snaith, *Science* **2013**, *6*, 1739.

[12]. M. Liu, M. B. Johnston, and H. J. Snaith. *Nature*, **2013**,*501*, 395.

[13]. H. Zhou, Q. Chen, G. Li, S. Luo, T. B. Song, H. S. Duan, Z. Hong, J. B. You,Y. Liu, and Y. Yang. *Science*, **2014,** *345*, 542.

[14]. W. S. Yang, J. H. Noh, N. J. Jeon, Y. C. Kim, S. Ryu, J. Seo, S. Il Seok, Sciencexpress, 2015,1.

[15].F. Hao, C. C. Stoumpos, D. H. Cao, R. P. H. Chang and M. G. Kanatzidis. *Nature photonics.* **2014**, *8*,489.

[16]. F. Hao, C. C. Stoumpos, R. P. H. Chang, M. G. Kanatzidis. *J. Am. Chem. Soc.* 2014, *136 (22)*,8094.

[17]. A. Mei, X. Li, L. Liu, Z. Ku, T. Liu, Y. Rong, M. Xu, M.Hu, J. Chen, Y. Yang, M. Grätzel, and H. Han, *Science*, **2014**, *345*,295.

[18]. Z. G. Xiao, C. Bi, Y. C. Shao, Q. F. Dong, Q. Wang, Y. B.Yuan, C. G. Wang, Y. Gao, and J.S. Huang, *Ener. Envir. Sci.* **2014,**7, 2619.

[19]. C. Bi, Y. C. Shao, Y. B. Yuan, Z. G. Xiao, C. G. Wang, Y. Gao, and J. S. Huang, *J. Mater. Chem.* **2014**, *A2*, 18508.

[20]. Q. Wang, Y. C. Shao, H. P. Xie, L. Lyu, X.L. Liu, Y. Gao, and J. S. Huang, *Appl. Phys. Lett.* **2014**,*105*, 163508.

[21]. X. Liu, C.G. Wang, L. Lyu, C. C. Wang, Z.G. Xiao, C. Bi, J.S. Huang, and Y. Gao, *Phys. Chem. Chem. Phys.* **2015,***17,* 896.

[22]. X. Liu, C.G. Wang, C.C. Wang, I. Irfan, and Y. Gao, *Org. Elec.* **2015**,*17*, 325.

[23].C.G. Wang, C.C. Wang, X.L. Liu, J. Kauppi, Y.C. Shao, Z.G. Xiao, C. Bi, J. S Huang, and Y. Gao, *Appl. Phys. Lett.* **2015,***106*, 111603.

[24].C.G. Wang, X.L. Liu, C.C. Wang, Z.G. Xiao, C. Bi, Y.C. Shao, J.S Huang, and Y. Gao, *J. Vac. Sci. Tech*. **2015,** B**33**, 032401.

[25]. J. Burschka, N. Pellet, S. J. Moon, R. Humphry-Baker, P. Gao, M. K. Nazeeruddin and M. Grätzel, *Nature*,**2013**, 499, 316.

[26]. J. B. You, Y. (Michael) Yang, Z. Hong, T. B. Song, L. Meng,Y. S. Liu, C.Y. Jiang, H. P. Zhou, W. H. Chang, G. Li, and Y. Yang. *Appl. Phys. Lett.*,**2014**,*105*, 183902.

[27]. G. Niu, W. Li, F. Meng, L.Wang, H. Dong and Y. Qiu, *J.Mater.Chem.A*, **2014**, *2*,705.

[28]. P. Schulz, E. Edri, S. Kirmayer, G. Hodes, D. Cahen and A. Kahn, *Energy Environ.sci.* **2014**, *7*,1377

[29]. Y. Yamada, T. Nakamura, M. Endo, A. Wakamiya and Y. Kanemitsu, *Appl.Phys. Express,* **2014**,*7*,032302.

[30]. I. G. Hill, A. J. Mäkinen, and Z. H. Kafafi, *J. Appl. Phys.* **2000,**88, 889.

[31]. A.V. Naumkin, A. Kraut-Vass, S. W. Gaarenstroom, and C. J. Powell, NIST X-ray Photoelectron Spectroscopy Database, http://srdata.nist.gov/xps/Default.aspx, accessed: June, 2000.

[32]. R. C. WEAST, handbook of chemistry and physics (General Chemical 51[st] edition), The Chemical Rubber Company, Germany, **1971**, section D

# Supporting Information

Shown in Figure S1 are the C 1s, I 3d, and Pb 4f XPS spectra of the $CH_3NH_3PbI_3$ thin film after oxygen exposure. There is no obvious change.

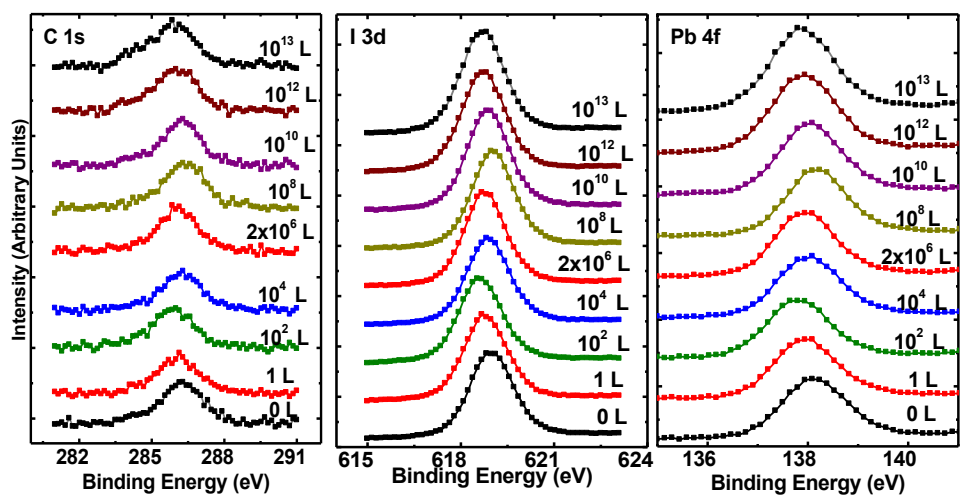

Figure S2 gives the C 1s, N 1s, I 3d, Pb 4f, O 1s and VBM of $CH_3NH_3PbI_3$ thin Films after air exposure. No obvious change was observed.

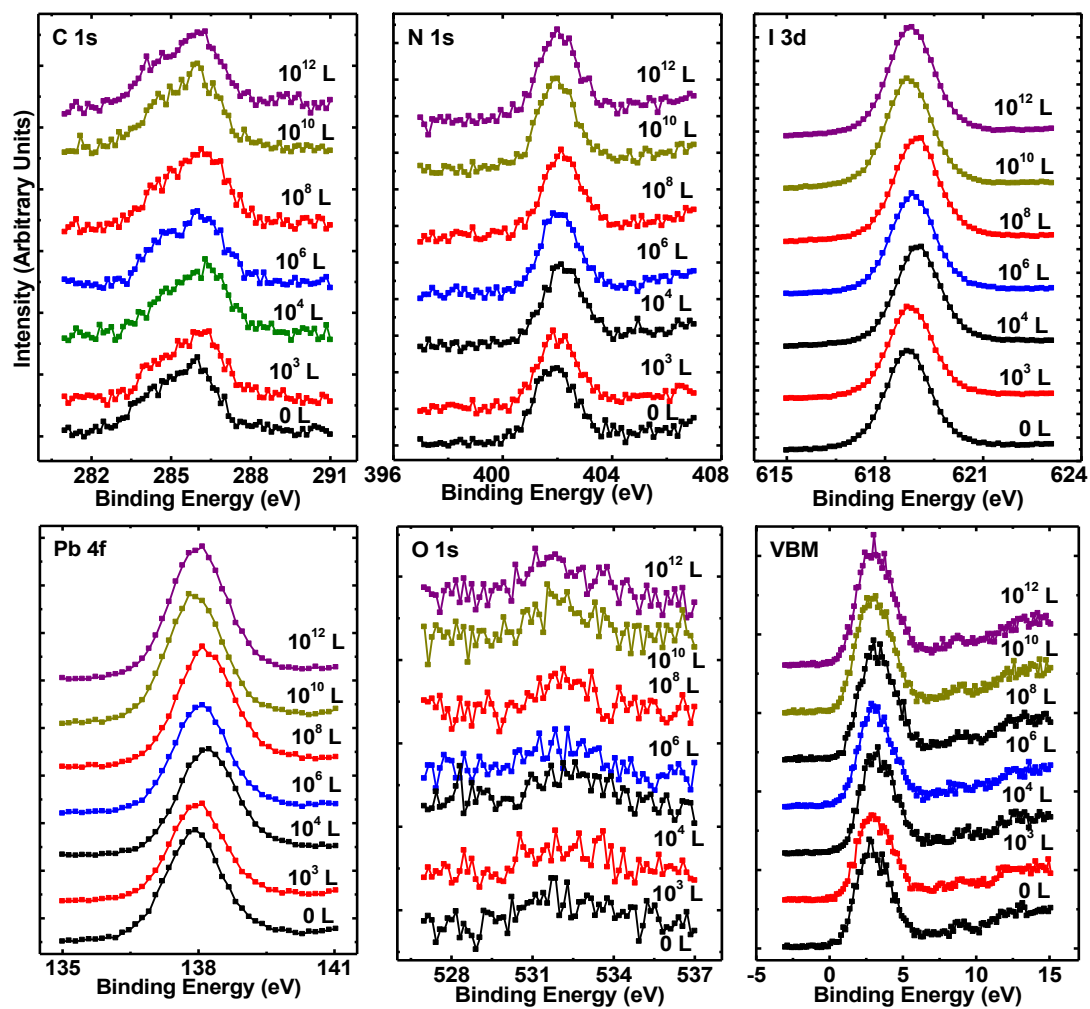